\documentclass[p1,12pt,preprint]{elsarticle}

\usepackage[margin=0.75in]{geometry}
\usepackage[utf8]{inputenc}
\usepackage[usenames,dvipsnames]{color}
\usepackage{amsmath}
\usepackage{bigints}
\usepackage{amsfonts}
\usepackage{mathrsfs}
\usepackage{amssymb}
\usepackage{ragged2e}
\usepackage{enumitem}
\usepackage{adjustbox}
\usepackage{array}
\usepackage{upgreek}
\usepackage[]{algorithm2e}
\usepackage{algorithmic}
\usepackage{graphicx}
\usepackage{epsfig}
\usepackage[T1]{fontenc}
\usepackage{tikz}
\usepackage{stackengine}
\usetikzlibrary{shapes.geometric}
\usetikzlibrary{arrows.meta,arrows}
\usepackage{color}
\usepackage{stmaryrd}
\usepackage{fancyhdr}
\usepackage{float}
\usepackage{rotating}
\usepackage{tikz}
\usepackage{pgfplots}
\usepackage{bm}
\usepackage{rotfloat}
\usepackage{multirow}
\usepackage{tabularx}
\usepackage{newunicodechar}
\usepackage{listings}
\usepackage[colorlinks=true]{hyperref}
\makeatletter
\AtBeginDocument{\def\@citecolor{Red}}
\AtBeginDocument{\def\@linkcolor{Cyan}}
\makeatother
\usepackage{caption}
\usepackage{subcaption}
\newunicodechar{​}{\ }
\usepackage{amsxtra, mathrsfs, upgreek, bm}
\usepackage{eucal}
\usepackage{textcomp,stmaryrd}
\DeclareMathAlphabet{\mathscr}{OT1}{pzc}{m}{it} 

\usepackage[mathletters]{ucs}
\usepackage{multicol}        
\usepackage[bottom]{footmisc}
\usepackage{latexsym}
\usepackage{booktabs}
\usepackage{siunitx}
\usepackage{arydshln}
\lstdefinestyle{mystyle}{
	backgroundcolor=\color{white},   
	commentstyle=\color{red},
	keywordstyle=\color{blue},
	numberstyle=\tiny\color{gray},
	stringstyle=\color{purple},
	basicstyle=\ttfamily\footnotesize,
	breakatwhitespace=false,         
	breaklines=true,                 
	captionpos=b,                    
	keepspaces=true,                 
	numbers=left,                    
	numbersep=5pt,                   
	showspaces=false,                
	showstringspaces=false,          
	showtabs=false,                  
	tabsize=2,                       
    frame=single,
	framesep=5pt,
	framerule=1pt,
	rulecolor=\color{black},
	escapechar=|,
	numbers=left,
	numberstyle=\tiny\color{gray},
	basicstyle=\ttfamily\small,
	}
\renewcommand{\arraystretch}{1.5}

\usepackage{titlesec}
\titleformat{\section}
{\normalfont\fontfamily{phv}\fontsize{14}{18}\bfseries}{\thesection}{1em}{}
\titleformat{\subsection}
{\normalfont\fontfamily{phv}\fontsize{12}{18}\bfseries}{\thesubsection}{1em}{}
\titleformat{\subsubsection}
{\normalfont\fontfamily{phv}\fontsize{10}{18}\bfseries\itshape}{\thesubsubsection}{1em}{}

\pgfplotsset{every tick label/.append style={font=\Large}}
\pgfplotsset{every axis/.append style={
		line width=2pt,
		tick style={line width=1.5pt}}}
\pgfplotsset{compat=1.4}

\newcommand{\ten}[1]{\bm{#1}}

\newcommand{\p}{\partial}
\newcommand{\pd}[2]{\frac{\p #1}{\p #2}}
\newcommand{\begeq}{\begin{equation}\begin{gathered}}
\newcommand{\eqend}{\end{gathered}\end{equation}}
\newcommand{\begal}{\begin{equation}\begin{aligned}}
\newcommand{\alend}{\end{aligned}\end{equation}}
\newcommand{\del}{\updelta }

\newcommand{\dd}{\, \mathrm d }

\journal{}

\begin{document}
	
	\begin{frontmatter}
		
		\title{A comparative analysis for different finite element types in strain-gradient elasticity simulations performed on Firedrake and FEniCS}
		
		
		\author[1]{B. Cagri Sarar}
		
		\author[2]{M. Erden Yildizdag}

		\author[3]{Francesco Fabbrocino}
		
		\author[4]{B. Emek Abali\corref{cor1}}

		\cortext[cor1]{Corresponding author}
		\ead{bilenemek@abali.org}
					
		\address[1]{International Research Center on Mathematics and Mechanics of Complex Systems, University of L'Aquila, L'Aquila, Italy}    
		\address[2]{Faculty of Naval Architecture and Ocean Engineering, Istanbul Technical University, 34469 Istanbul, Turkey}
		\address[3]{Department of Engineering, Telematic University Pegaso, 80143 Napoli, Italy}
		\address[4]{Department of Materials Science and Engineering, Division of Applied Mechanics, Uppsala University,Box 35, 751 03 Uppsala, Sweden}

\begin{abstract}
The layer-upon-layer approach in additive manufacturing, open or closed cells in polymeric or metallic foams involve an intrinsic microstructure tailored to the underlying applications. Homogenization of such architectured materials creates metamaterials modeled by higher-gradient models, specifically when the microstructure's characteristic length is comparable to the length scale of the structure. In this study, we conduct a comparative analysis of various finite elements methods for solving problems in strain-gradient elasticity. We employ open-source packages from Firedrake and FEniCS. Different finite element formulations are tested: we implement Lagrange, Argyris, Hermite elements, a Hu--Washizu type (mixed FEM) formulation, as well as isogeometric analysis with Non-Uniform Rational B-Splines (NURBS). For the numerical study, we investigate one- and two-dimensional problems discussed in the literature of strain-gradient modeling. Among the examined formulations, Argyris and mixed FEM demonstrate superior accuracy, whereas Hermite and IGA lack of convergence behavior. Displacements predicted by Hermite elements also differ noticeably in the 1-D case. All developed codes are open-access to encourage research in Finite Element Method (FEM) based computation of generalized continua.
\end{abstract}

		
\end{frontmatter}

\section{Introduction}

Owing to its innovative process, additive manufacturing (AM) facilitates the fabrication of extraordinary structures, thereby providing engineers an opportunity to envision and explore diverse perspectives in design and manufacturing \cite{colosimo2022complex, zyla20243d, barchiesi2024complex}. In particular, the layer-by-layer production technique offered by AM enables the creation of intricate internal patterns, controlling and enhancing the functionality of the manufactured components (for example,  see recent studies \cite{bi2022continuous,turco2024harnessing, dos2024deep}). Therefore, the ability to adjust infill patterns \cite{ambati2022effect, turco2024long, rahmatabadi20244d} and modify infill ratios \cite{aydin2022strain,afshar2023nonlinear,casalotti2024optimization} has become pivotal in shaping the internal structural design of multi-scale structures \cite{aydin2022investigating,torisaki2023shape, allena2023model}. This functionality not only enables engineers to fine-tune the mechanical properties and performance characteristics of manufactured components; but also empowers them to optimize material usage and reduce manufacturing costs.

Designing and fabricating complex materials using AM clearly necessitates novel mathematical models for assessing the overall behavior under various conditions. As classical mathematical models fall short in accurately predicting the behavior of multi-scale structures, alternative advanced modeling techniques, which require proper generalization of classical elasticity theory, are currently drawing great deal of interest (see recent applications in \cite{giorgio2024second,yilmaz2024emergence, stilz2022continuum}). Particularly, micropolar \cite{chroscielewski2020rotational, tian2023thermodynamics, vilchevskaya2023modeling},  micromorphic \cite{misra2021identification,seppecher2022asymptotic, sarhil2024computational}, Cosserat \cite{giorgio2020biot,giorgio2022experimental,giorgio2023geometrically}, strain-gradient \cite{abali2015strain,yang2021verification,eremeyev2021nonlinear, giorgio2021lattice, rezaei2024procedure} continua are extensively investigated to model complex problems providing efficient numerical simulations. Notably, to establish such models, variational methods \cite{lekszycki2020variational,placidi2020variational, misra2020variational} are quite powerful, providing a systematic approach to study higher gradient theories \cite{maksimov2021two, ciallella2023deformation, sessa2024implicit, alibert2003truss}. Basically, these variational frameworks involve defining an appropriate action functional, which encapsulates the physical properties and constraints of the system under consideration. By processing the action functional concerning the relevant fields, such as displacement, strain, or other relevant entities, one may derive the Euler--Lagrange equations that govern the problem under investigation \cite{dell2022second,dell20205}. Specifically, in the context of strain-gradient modeling, the action functional depends on not only the strain but also its gradient. This extension may be motivated by an upscaling from a discrete system with non-local connection \cite{065} or continuum setting starting from Cauchy continuum \cite{075}. 

In addition to modeling, effective numerical simulation of such complex models is also of great interest to provide reliable predictions. In this study, we present an in-depth comparative analysis for the numerical simulation of strain-gradient models, considering different finite element formulations. To this end, two benchmark problems are investigated that are available in the literature, by utilizing packages from FEniCS and Firedrake, open-source finite element computing platforms. In each case, the problem is simulated using various shape functions, including Lagrange, Argyris, Hermite elements, a mixed FE formulation, and NURBS-based isogeometric analysis (IGA). We investigate FE solution of a two-dimensional simple shear problem discussed in \cite{shekarchizadeh2022benchmark} and a one-dimensional pull-out problem numerically studied in \cite{rezaei2024strain, rezaei2022solution}. For the FE analysis, a discrete approximation of continuous functions are used in order to solve a variational formulation that is challenging to adequately represent by a suitable function space. Hence, there are different suggestions in the literature. Firedrake and FEniCS are using modern techniques with high level scripting to solve any formulation that automates the numerical solution of field equations with the chosen element type and formulation. Specifically herein, $C^1$ continuous elements \cite{greco2024objective}, and isogeometric analysis (IGA) as function spaces are rather simple to utilize by leveraging the domain-specific language of the FEniCS project \cite{rathgeber2016firedrake}. FE method is based on a compact support with elements only locally effective such that a monotonous convergence is guaranteed. Yet higher order formulations are often defined within the whole domain (sometimes called patch) such that reliability and error estimation may be in danger. One example of nonlocal formulation is the so-called isogeometric analysis (IGA), where field equations are mapped to a parametric reference domain, and solutions are approximated using function spaces constructed from linear combinations of finite element basis functions over that domain \cite{hughes2005isogeometric}. IGA has been extensively applied into different fields (for example, see \cite{cazzani2016isogeometric,schulte2020isogeometric,yildizdag2022isogeometric}). In this work, we use the library called tIGAr \cite{kamensky2019tigar} available within the open-source finite element automation software FEniCS. It employs a global version of Bézier extraction to integrate FEniCS's capabilities into IGA workflows. Furthermore, with the aid of Firedrake platform, we exploit Argyris and Hermite, which are higher-order elements. Another possibility is to increase the number of field equations such that the smoothness conditions are reduced leading to a mixed formulation.

The rest of the study is as follows. In Section 2, the adapted problems and their strain-gradient models are summarized. In Section 3, the conducted comparative analysis is demonstrated and discussed in detail, assessing the different FE formulations. Finally, conclusions are drawn in Section 4.

\section{Method of solution in strain-gradient elasticity}

The strain-gradient model \cite{030} is adapted in the numerical simulations and summarized in this part. Following the variational framework presented in \cite{abali2019revealing}, we consider the Lagrange function depending on displacement and its first and second gradients in space and time, as follows:
\begin{align}
\mathcal{L} =  - w + \rho_0 (f_i u_i + \ell_{ij} u_{i,j})
\end{align}
where we understand Einstein summation convention over the repeated indices, $i,j,k,\dots \in [1,2,3]$ in three-dimensional space,  $\rho_0$ is the mass density of the material, and $\ten u$ is the displacement, defined by difference between the current position, $\ten x$, and the reference position, $\ten X$. The kinetic energy term in Lagrangian density is omitted since the problem is static; interested readers may refer to \cite{abali2019revealing} for further details.
The first term, $w$, is the deformation energy density, which is a function of the linearized Green--Lagrange strain and its gradient, $w = \hat{w} (\ten\varepsilon, \nabla\ten\varepsilon)$. The last term in the Lagrangian density represents the energy density due to the volumetric effects, including a body force density, $\ten f$, and a supply term, $\ten\ell$, which is accounted for work through displacement gradient. The linearized strain tensor and its gradient are defined by
\begin{align}
	\varepsilon_{ij} = \frac{1}{2} (u_{i,j} + u_{j,i}), \quad \quad \varepsilon_{ij,k} = \frac{1}{2} (u_{i,jk} + u_{j,ik})
\end{align}
where 
\begin{align}
u_{i,j} = \frac{\partial u_i}{\partial X_j}, \quad \quad u_{i,jk} = \frac{\partial^2 u_i}{\partial X_k \partial X_j} \ .
\end{align}
The deformation energy density is given by
\begin{align}
w = \frac{1}{2} \varepsilon_{ij} C_{ijkl} \varepsilon_{kl}  + \frac12 \varepsilon_{ij,k} D_{ijklmn} \varepsilon_{lm,n} + \varepsilon_{ij} G_{ijklm} \varepsilon_{kl,m} \ ,
\end{align}
where the rank-4 tensor, \( C_{ijkl} \), corresponds to first-gradient (strain related) elastic properties. For isotropic materials, it reads 
\begin{align}
C_{ijkl} = c_1 \delta_{ij} \delta_{kl} + c_2 (\delta_{ik} \delta_{jl} + \delta_{il} \delta_{jk}) \ .
\end{align}
The rank-6 tensor, \( D_{ijklmn} \), represents second-gradient (strain gradient related) elastic properties, again for isotropic materials,
\begal
D_{ijklmn} = & \, c_3 (\delta_{ij} \delta_{kl} \delta_{mn} + \delta_{in} \delta_{jk} \delta_{lm} + \delta_{ij} \delta_{km} \delta_{ln} + \delta_{ik} \delta_{jn} \delta_{lm}) + c_4 \delta_{ij} \delta_{kn} \delta_{ml} \\
		& \ + c_5 (\delta_{ik} \delta_{jl} \delta_{mn} + \delta_{im} \delta_{jk} \delta_{ln} + \delta_{ik} \delta_{jm} \delta_{ln} + \delta_{il} \delta_{jk} \delta_{mn})+  c_6 (\delta_{il} \delta_{jm} \delta_{kn} + \delta_{im} \delta_{jl} \delta_{kn})  \\
		& \ + c_7 (\delta_{il} \delta_{jn} \delta_{mk} + \delta_{im} \delta_{jn} \delta_{lk} + \delta_{in} \delta_{jl} \delta_{km} + \delta_{in} \delta_{jm} \delta_{kl}) .
\alend
Strain gradient is coupled to strain via the rank-5 tensor, (\(G_{ijklm}\) = 0), which vanishes for a centro-symmetric microstructure as studied herein. Also, $c_1$ and $c_2$ are first-gradient constitutive parameters while $c_3, c_4, c_5, c_6,$ and $c_7$ are strain-gradient constitutive parameters.
 
Next, the following action functional is postulated in space (3-D) and time spanning an infinitesimal element, $\dd\Sigma=\dd V \dd t$, for volume and time; as well as an infinitesimal element, $\dd \Gamma=\dd A \dd t$, for surface and time; and herein, $\dd\Pi=\dd\ell \dd t$, for line (edge) and time, as follows:
\begin{align}
\mathcal{A} = \int_{\Omega} \mathcal{L} \, d\Sigma + \int_{\partial\Omega} W_s \, d\Gamma + \int_{\partial\partial\Omega} W_e \, d\Pi
\end{align}
where $W_\text{s}$ and $W_\text{e}$ are energy densities defined, respectively, over the surface ${\partial\Omega}$ and the edge ${\partial\partial\Omega}$ of the domain of interest ${\Omega}$. These so-called \textsc{Neumann} boundaries are given. We neglect the inertial terms in this study and compute cases in statics such that the time integration drops. Then, the variation of the energy reads
\begal
\del \mathcal{A} = \int_{\Omega} & \left( \rho_0 f_i \del u_i + \rho_0 \ell_{ij} \del u_{i,j} - \frac{\partial{w}}{\partial{u_{i,j}}} \del u_{i,j}  - \frac{\partial{w}}{\partial{u_{i,jk}}} \del u_{i,jk} \right) \dd V \\
& + \int_{\partial\Omega} \bigg( \pd{W_\text{s}}{u_i} \del u_i + \pd{W_\text{s}}{u_{i,j}} \del u_{i,j} \bigg)  \, \dd A + \int_{\partial\partial\Omega} \pd{W_\text{e}}{u_i} \del u_i \, \dd\ell \ .
\label{eq:var_general}
\alend
A vanishing action leads to the solution according to \textsc{Noether}'s theorem \cite{abali2023energy}. The test function, $\del \ten u$, is arbitrary and chosen to vanish on the \textsc{Dirichlet} boundaries, where the primitive variables are known. Both $\ten u$ and $\del\ten u$ are are from same space known as the \textsc{Galerkin} approach. The derivative of the stored energy density with respect to the first gradient of displacement and the second gradient of displacement are derived as follows
\begin{align}
\frac{\partial w}{\partial u_{i,j}} &= \frac{\partial w}{\partial \varepsilon_{kl}} \frac{\partial \varepsilon_{kl}} {\partial u_{i,j}} = \frac{\partial w}{\partial \varepsilon_{kl}} \frac{1}{2} (\delta_{ki} \delta_{lj} + \delta_{kj} \delta_{li}) = C_{ijkl} \varepsilon_{kl}, \label{eq:var_first_order} \\
\frac{\partial w}{\partial u_{i,jk}} &= \frac{\partial w}{\partial \varepsilon_{lm,n}} \frac{\partial \varepsilon_{lm,n}}{\partial u_{i,jk}} = \frac{\partial w}{\partial \varepsilon_{lm,n}} \frac{1}{2} (\delta_{li} \delta_{mj} \delta_{nk} + \delta_{mi} \delta_{lj} \delta_{nk}) = D_{ijklmn} \varepsilon_{lm,n}, \label{eq:var_second_order}
\end{align}
by considering the symmetry of the stiffness tensors (\(C_{ijkl} = C_{jikl} = C_{klij} = C_{klji} \)) and (\( D_{lmnijk} = D_{lmnjik} = D_{ijklmn} = D_{jiklmn}\)).

In order to perform simulations with different finite element formulations, Eq. \eqref{eq:var_general} is utilized in FEniCS and Firedrake. Neglecting inertial terms, body forces, and work on edge and surface due to second gradient terms, Eq. \eqref{eq:var_general} takes the following form:
\begin{align}
 \int_{\Omega} \left( \del u_{i,j} C_{ijkl} \varepsilon_{kl} 
	+   \del u_{i,jk} D_{ijklmn}  \varepsilon_{lm,n} \right) \dd V 
	= \int_{\partial\Omega} \hat t_i \del u_i \dd A \ ,
	\label{traction}
\end{align}
where $\hat{\ten t}$ is the traction given on boundaries.

One possible approach is to use sufficiently smooth elements for $\ten u$. In this case, for applying boundary conditions (i.e. prescribed displacement, prescribed displacement normal gradient, and applied tractions on both boundary and edge), the penalty method is utilized in the simulations. To properly enforce the boundary conditions while using IGA, Argyris and Hermite elements, Eq. \eqref{traction} results in
\begal
 \ \int_{\Omega} \left( \del u_{i,j} C_{ijkl} \varepsilon_{kl} +   \del u_{ij,k} D_{ijklmn}  \varepsilon_{lm,n} \right) \dd V 
	= \int_{\partial\Omega_\text{N}}  \hat t_i \del u_i \dd A  + \int_{\partial\Omega_\text{D}} K (u_i-\hat u_i) \del u_i + K (\del u_i-\hat {\delta u_i}) \del u_i \dd A \ ,
\alend
with given Neumann boundaries, $\hat{\ten t}$, and Dirichlet boundaries, $\hat{\ten u}$, $\hat{\ten \delta \ten u}$. The implementation of Dirichlet boundaries is employed in the integral form by means of a large number, $K$, also called penalty factor. This fact is needed owing to higher continuity of the element at the domain boundaries.

Another possibility is to introduce the mixed FE formulation \cite{shekarchizadeh2022benchmark} with an additional unknown, $\mathit{g}_{ij}$, imposing $\mathit{u}_{i,j} = \mathit{g}_{ij}$, through Lagrange multipliers as follows:
\begeq
\int_{\Omega} \Big( \del u_{i,j} C_{ijkl} \varepsilon_{kl} + \del g_{ij,k} D_{ijklmn}  \varepsilon_{lm,n} + L_{ij} (\del g_{ij} - \del u_{i,j} ) + (g_{ij} - u_{i,j} )\del L_{ij}\Big) \, \dd V
= \int_{\partial\Omega_\text{N}} \hat t_i \del u_i \, \dd A \ , 
\eqend
where
\begeq
\varepsilon_{ij} = \frac12 ( u_{i,j} + u_{j,i} ) \ , \ 
\varepsilon_{ij,k} = \frac12 ( g_{ij,k} + g_{ji,k} ) \ .
\eqend
In this version, Dirichlet boundaries are implemented directly by taking them out of the matrices after assembly in the finite element method.

\subsection{2-D simple shear problem}

\begin{figure}[H]
	\centering
	\begin{subfigure}{0.49\textwidth}
		\centering
		\includegraphics[width=1\linewidth]{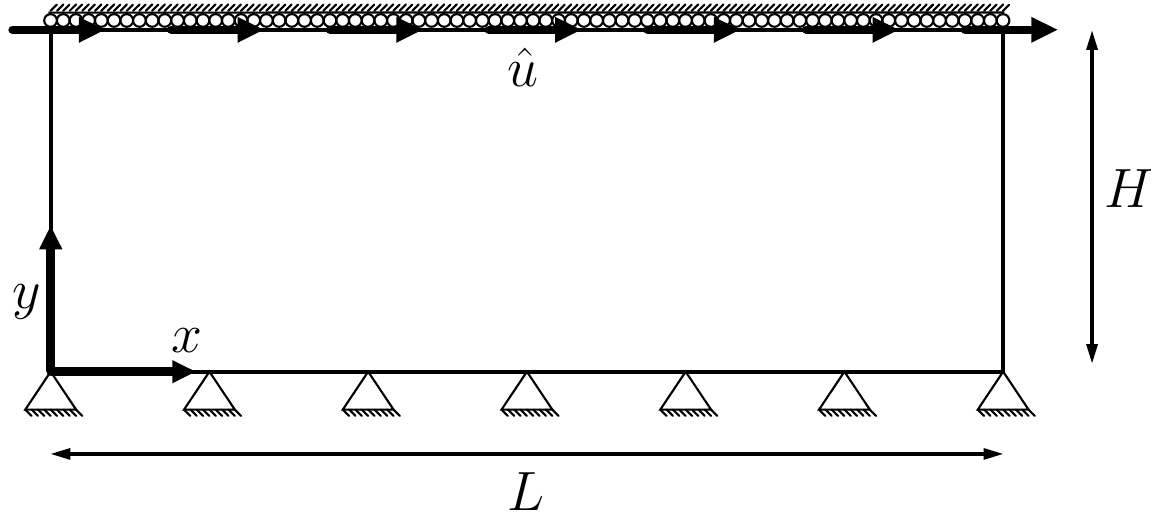}
		\caption{Prescribed displacement, case D}
		\label{shear1}
	\end{subfigure}
	\begin{subfigure}{0.49\textwidth}
		\centering
		\vspace{20mm} 
		\includegraphics[width=1\linewidth]{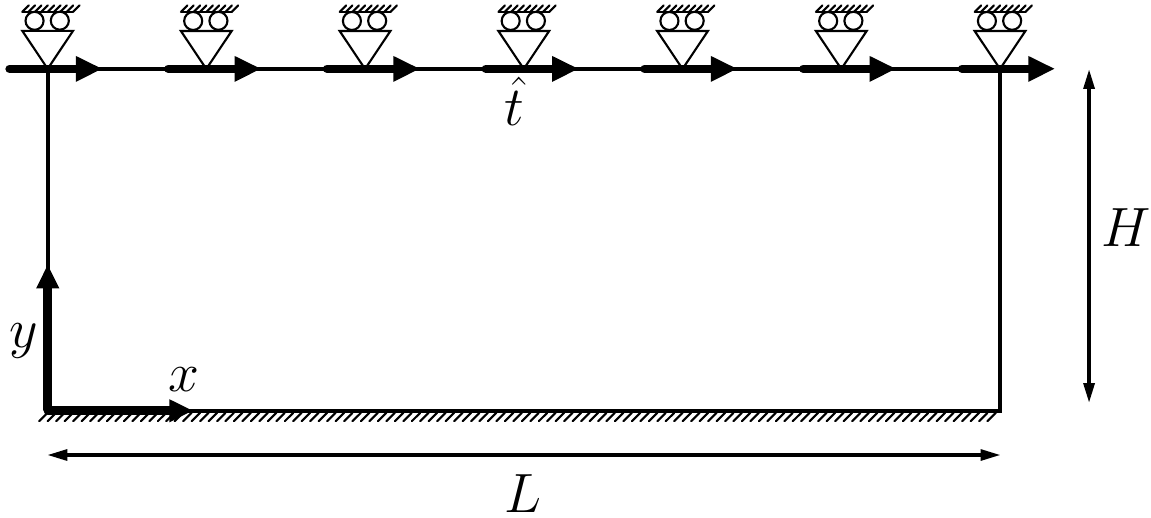}
		\caption{Applied traction, case T}
		\label{shear2}
	\end{subfigure}
	\caption{2-D simple shear problem for solving by the strain-gradient formulation with two separate boundary conditions.}
	\label{shear}
\end{figure}

The first strain-gradient model examined in this study is the 2-D simple shear of a plate of length $L$ and height $H$, which was investigated both numerically and analytically by Shekarchizadeh et al. \cite{shekarchizadeh2022benchmark}. The general analytical solution in closed form is provided below,
\begin{align}
	u = u_x(y) = q_1 + q_2 y + q_3 \sinh \Big( \frac{y}{r} \Big) + q_4 \cosh \Big(\frac{y}{r}\Big)
\end{align}
where $r = \sqrt{\frac{(c_5 + c_6 + c_7)}{c_2}}$ with integration constants $q_1$, $q_2$, $q_3$, $q_4$. We study 2 cases:
\begin{itemize}
\item
case D: the prescribed displacement ($\hat{\ten u}$):
\begeq
	q_1 = 0, 
	\quad
	q_2 = \frac{\hat{u} \cosh (\frac{H}{r})}{H \cosh (\frac{H}{r}) - r \sinh (\frac{H}{r})  },
	\quad
	q_3 = \frac{\hat{u} r}{r \sinh (\frac{H}{r}) - H \cosh (\frac{H}{r})  },
	\quad
	q_4 = 0
\eqend
\item
case T: the applied traction (as a vector \(\mathbf{T} = (0, \hat{t})\) at \(y = H\)):

\begeq
	q_1 = \frac{-r^3 \hat{t} \sinh (\frac{H}{r})}{s}, 
	\quad
	q_2 = \frac{r^2 \hat{t} \cosh (\frac{H}{r})}{s},
	\quad
	q_3 = \frac{-r^3 \hat{t} \cosh (\frac{H}{r})}{s},
	\quad
	q_4 = \frac{r^3 \hat{t} \sinh (\frac{H}{r})}{s}
\eqend
where 
\begeq
	s = c_2 r^2 \left( \cosh\left(\frac{H}{r}\right)- 1 \right) + c_5 + c_6 + c_7 \ .
\eqend
\end{itemize}
As schematically shown in Fig.\,\ref{shear}, for case D in Fig.\,\ref{shear1} and case T in Fig.\,\ref{shear2}, the difference between these cases is obvious at the bottom edge. In case D, the bottom edge may rotate, while in case T, it is clamped.
%
%

For the prescribed displacement, case D, a displacement of $\hat{\ten u}$ is applied at the top edge by a rail such that the displacement gradient is set to zero in order to illuminate the effects of strain-gradient terms. While the bottom edge prevents a translation but not rotation. For this case, the weak form reads
\begal
	\int_{\Omega} & \Big( \del u_{i,j} C_{ijkl} \varepsilon_{kl} + \del u_{ij,k} D_{ijklmn}  \varepsilon_{lm,n} \Big) \dd V = 
		\int_{\partial\Omega_\text{bot}}  K (u_i - \hat u_i^\text{bot} ) \del u_i \dd A
	\\ &
	 + \int_{\partial\Omega_\text{top}} \Big( K (u_i - \hat u^\text{top}_i ) \del u_i + K ( u_{i,j} - \hat g^\text{top}_{ij} ) \del u_{i,j} \Big) \dd A 
\alend
where all terms with hat, $\hat{\ten u}^\text{bot}=0$, $\hat{\ten u}^\text{top}$, $\hat{\ten g}^\text{top}=0$, are given to prescribe displacement. 

For the applied traction, case T, a traction, $\hat{\ten t}$ is applied at the top edge, while the displacement and displacement gradient are fixed to zero at the bottom edge, 
\begal
	\int_{\Omega} & \Big( \del u_{i,j} C_{ijkl} \varepsilon_{kl} + \del u_{ij,k} D_{ijklmn}  \varepsilon_{lm,n} \Big) \dd V = 
	 \int_{\partial\Omega_\text{top}} \hat t_i \del u_i \dd A  \\
	& + \int_{\partial\Omega_\text{top}} K (u_1-\hat u^\text{top}_1 ) \del u_1  \dd A 
	+ \int_{\partial\Omega_\text{bot}} \Big( K ( u_i-\hat u^\text{bot}_i ) \del u_i + K (u_{i,j} - \hat g^\text{bot}_{ij} ) \del u_{i,j} \Big)
	\dd A 
\alend
although written explicitly, $\hat u^\text{top}_1=0$, $\hat{\ten u}^\text{bot}=0$, $\hat{\ten g}^\text{bot}=0$, setting all components zero ensure no deformation and no rotation along the bottom edge.

Importantly, periodic boundary conditions are imposed to the lateral boundaries of the plate, effectively rendering the plate's length negligible.

\subsection{1-D pull-out problem}

\begin{figure}[H]
	\centering
	\includegraphics[width=0.8\textwidth]{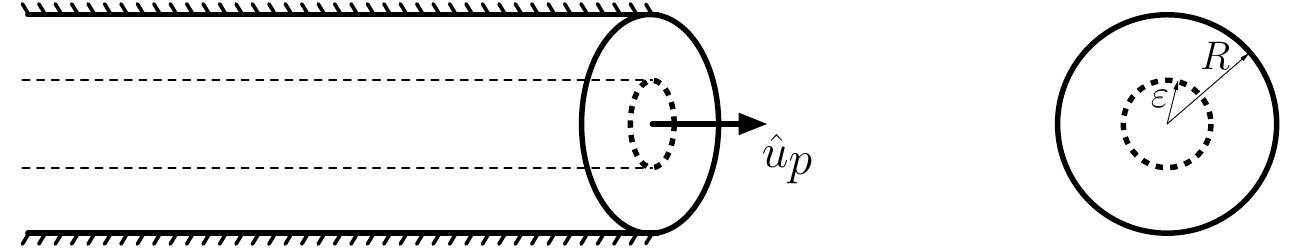}
	\caption{The pull-out problem, a prescribed displacement, $\hat u_p$, pulls the rebar from the cylindrical bulk of radius $R$ clamped around the surface (skin) area.}
	\label{fig:pull-out}
\end{figure}

The pull-out problem is the second strain-gradient model examined in this study by following Rezaei et al. \cite{rezaei2024strain}. As shown in Fig.\,\ref{fig:pull-out}, the pull-out test is an extraction of a rebar (reinforced rod) from a cylindrical block of radius $R$. By following Rezaei et al. \cite{rezaei2024strain}, the axisymmetric 3-D problem is simplified to formulate a 1-D model. Here, the assumptions are: (i) external surface of the cylindrical block is fixed; (ii) cylindrical block is infinitely long to neglect boundary effects and have displacement field uniform in circumferential direction; (iii) the imposed displacement applied to rigid bar (${\hat{u}}_p$) is axially directed and the radius of the rigid bar, $\varepsilon$, goes to zero.
For a solution, the system is reduced to a 1-D formulation by utilizing cylindrical polar coordinates, (\(r,\theta,z\)), such that the quadratic deformation energy reads in Green--Lagrange strain and their gradients,
\begal
w(u_z, u'_z, u''_z ) = & \frac{\lambda}{8} (u_z')^4 + \mu \Big( \frac{(u_z')^2}{2} + \frac{(u_z')^4}{4} \Big)  \\
		& + \frac{a_1}{2r} \Big( (u_z')^2 u_z'' + (u_z' + 2ru_z'') \Big) + 2a_2 (u_z')^2 (u_z'')^2  \\
		&  + \frac{a_3}{4r^2} \Big( (u_z')^2 (1 + 4r^2 (u_z'')^2) + r^2 (u_z'')^2 + 4r (u_z')^3 u_z'' + 2r u_z' u_z'' + (u_z')^4 \Big)  \\
		& + \frac{a_4}{2r^2} \Big( (u_z')^2 (2r^2 (u_z'')^2 + 1) + r^2 (u_z'')^2 + (u_z')^4 \Big)  \\
		&  + \frac{a_5}{4r^2} \Big( (u_z')^2 (4r^2 (u_z'')^2 + 1) + r^2 (u_z'')^2 + (u_z')^4 \Big)
\alend
where $c_1$ and $c_2$ are the first-gradient parameters (i.e., Lame parameters), and $c_3, c_4, c_5, c_6$, and $c_7$ are second-gradient parameters. Also, $u_z$ represents the axial displacement. Considering the analogous variational formulation, we obtain
\begeq
\del \mathcal{A} = 0 \int_{\Omega} \Big( \frac{\partial w}{\partial u_z} \delta u_z + \frac{\partial w}{\partial u'_z} \delta u'_z + \frac{\partial w}{\partial u''_z} \delta u''_z \Big) \dd V \ .
\eqend
By calculating each term,
\begal
	\frac{\partial w}{\partial u_z} &= 0, \\
	\frac{\partial w}{\partial u'_z} &= \frac{\lambda}{2} \left( (u_z')^3  \right) + \mu \left( (u_z')  + (u_z')^3  \right)  + \frac{a_1}{2z} \left( 3(u_z')^2 u'' + 4z (u_z') (u'')^2  \right) + 2a_2 \left( 2(u_z') (u'')^2 \right)  \\ 
	& +\frac{a_3}{4z^2} \biggl( 2(u_z') + 8z^2 (u_z') u'' (u_z')^2 + 12z (u_z')^2 u'' (u_z') + 2z u'' + 4(u_z')^3 \biggr)  \\
	& + \frac{a_4}{2z^2} \left( 4z^2 (u_z') (u'')^2 +  2(u_z') + 4(u_z')^3 \right) + \frac{a_5}{4z^2} \left( 8z^2 (u_z') (u'')^2  +  2(u_z') + 4(u_z')^3 \right), \\
	\frac{\partial w}{\partial u''_z} &= \frac{a_1}{2z} \left( (u_z')^3 + 4z u'' (u_z')^2  \right) + 2a_2 \left( 2u'' (u_z')^2\right)  \\
	& + \frac{a_3}{4z^2} \biggl( 8z^2 (u_z')^2 u'' + 2z^2 u'' \delta u_z'' + 4z (u_z')^3 + 2z (u_z') + 4(u_z')^3 \biggr)   \\
	& + \frac{a_4}{2z^2} \left( 4z^2 (u_z')^2 u'' +  + 2z^2 u'' \right) + \frac{a_5}{4z^2} \left( 8z^2 (u_z')^2 u'' +  + 2z^2 u''  \right). 
\alend
We stress that the adapted 1-D strain-gradient model is non-linear without an analytical solution as opposed to the 2-D strain-gradient model presented for the simple shear problem with the aforementioned analytical solution.

\section{Numerical results and discussion}

For the numerical simulations of the 2-D strain-gradient model, we consider a plate of length $L=1.5$ mm and height $H = 0.5$ mm. The constitutive parameters adapted in the simulations are listed in Table \ref{tab:material_params2}, considering a material of Young's modulus E = 400 MPa and Poisson's ratio $\nu$ = 0.49. All the constitutive parameters, including those for the strain-gradient elasticity contribution, are based on the granular micromechanics modeling by Barchiesi et al. \cite{barchiesi2021granular}, given by,
\begal
	& c_1 = \frac{E \nu}{(1+ \nu)(1-2\nu)}  
	\\ 
	& c_2 = \frac{E}{2(1+\nu)} 
	\\ 
	\label{constitutive}
	& c_3 = c_4 = \frac{l^2}{112} \lambda \\ 
	& c_5 = c_7 = \frac{l^2}{1120} (7\mu + 3\lambda)
	\\ 
	& c_6 = \frac{l^2}{1120} (7\mu - 4\lambda) \ ,
\alend
where the characteristic length, $l = 0.1$ for case D and $l = 0.2$ for case T, represents the size of microstructural interactions.

The parameters $c_1$ and $c_2$ compensate the negative values $c_3$, $c_4$, $c_5$, $c_6$, $c_7$, we emphasize that the presence of negative values is a known fact \cite{thbaut2024effective} and still resembles a positive energy, thereby guaranteeing a unique solution. This aspect is elaborately discussed in the context of positive definiteness in strain-gradient theory \cite{nazarenko2021positive,dell2009generalized,eremeyev2020well}.
 
\begin{figure}[H]
	\centering
	\begin{subfigure}{0.49\textwidth}
		\centering
		\includegraphics[width=0.95\linewidth]{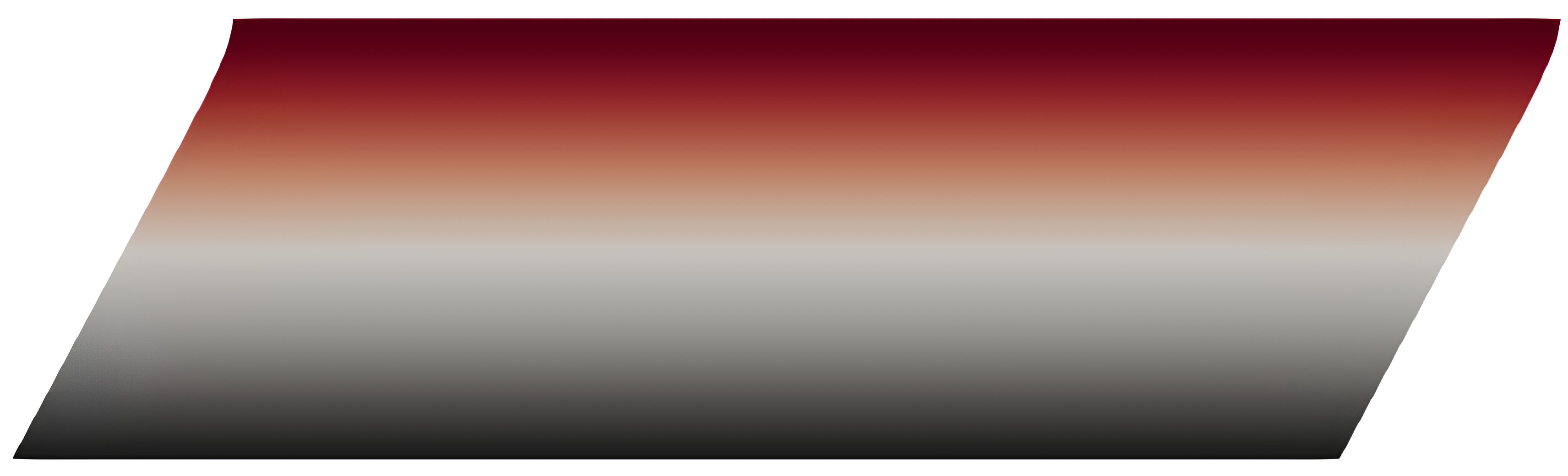}
	\end{subfigure}
	\begin{subfigure}{0.49\textwidth}
	\centering
	\includegraphics[width=0.9\linewidth]{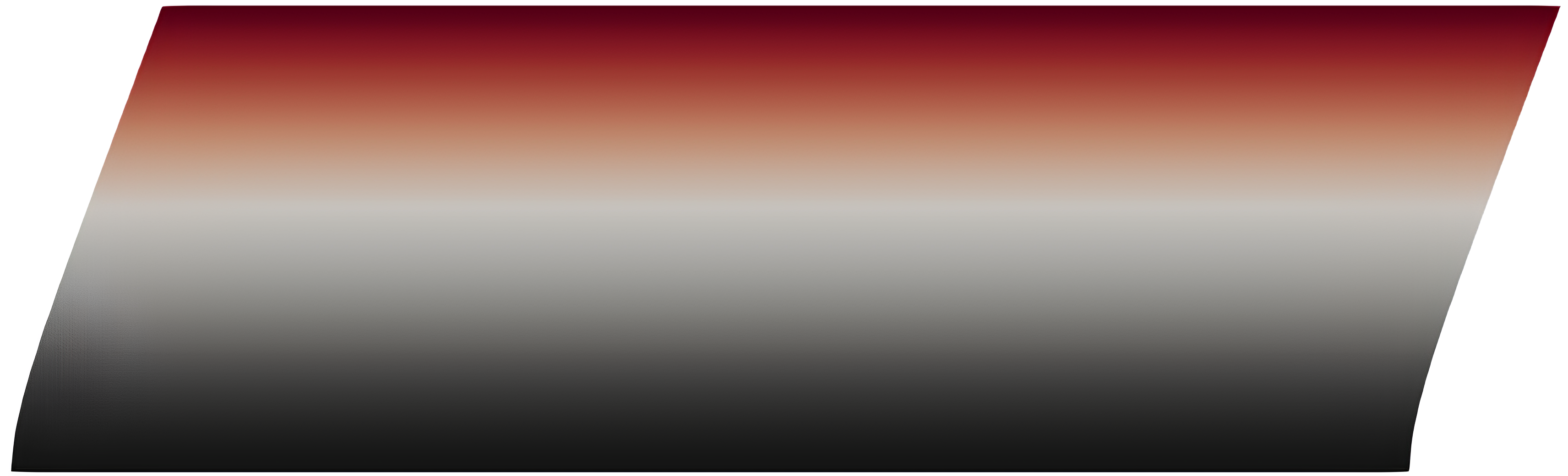}
	\end{subfigure}

	\begin{subfigure}{0.49\textwidth}
		\centering
		\includegraphics[width=0.98\linewidth]{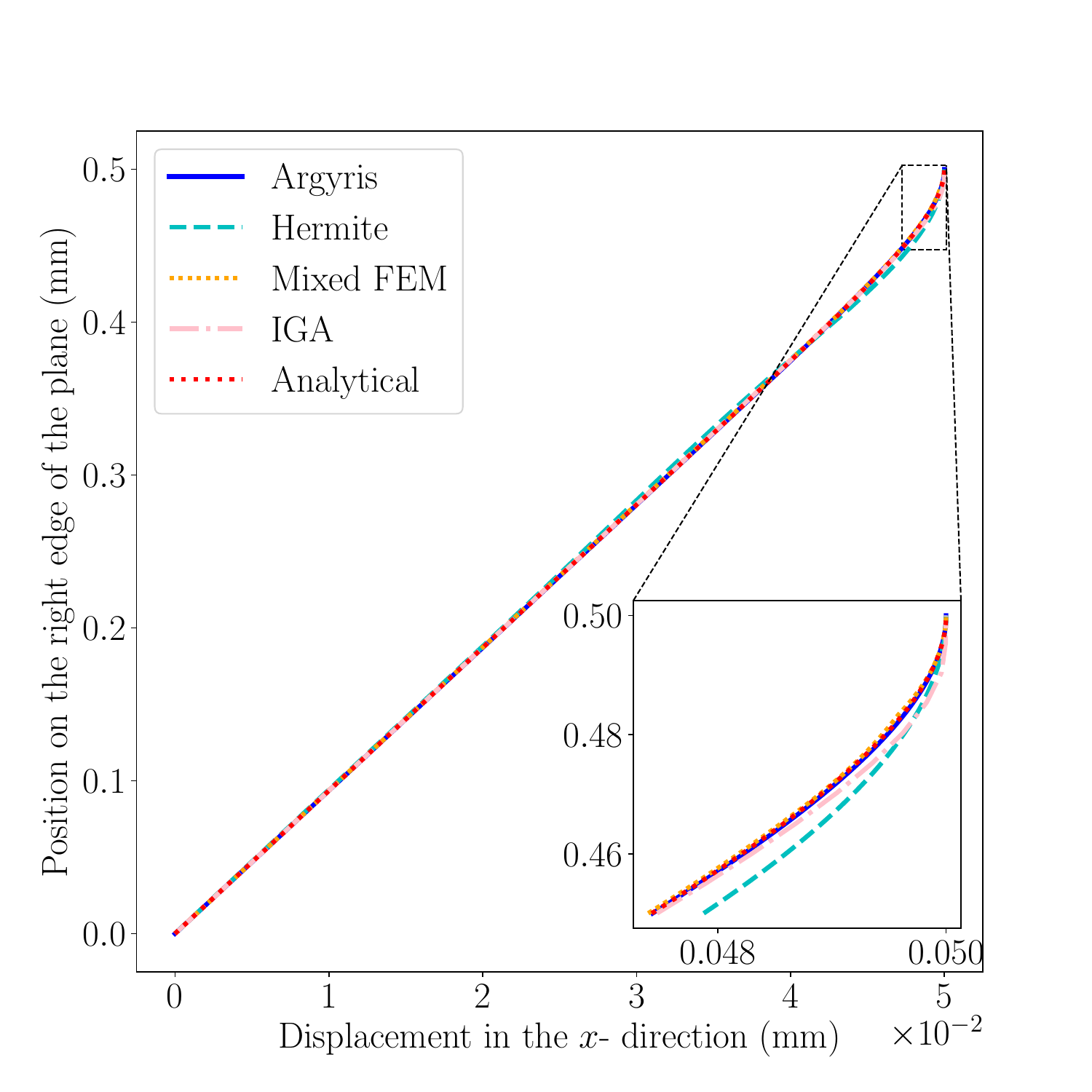}
		\caption{Prescribed displacement case.}
		\label{shear_sim1}
	\end{subfigure}
	\begin{subfigure}{0.49\textwidth}
		\centering
		\includegraphics[width=0.98\linewidth]{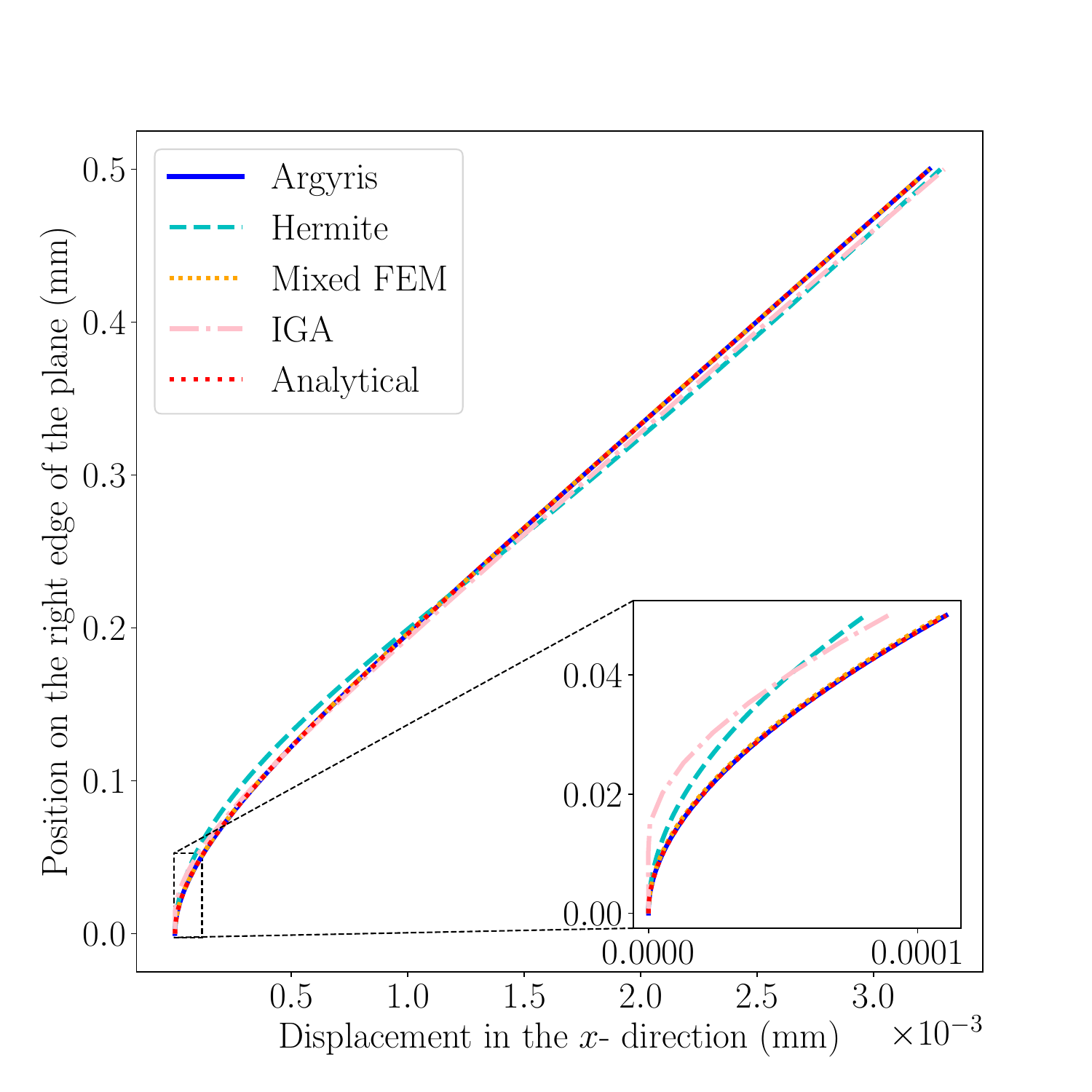}
		\caption{Applied traction case.}
		\label{shear_sim2}
	\end{subfigure}
	\caption{Comparison of the converged results for the simple shear problem.}
	\label{shear_sim}
\end{figure}

\begin{table}[H]
	\centering
	\caption{Constitutive parameters adapted for the simple shear problem.}
	\label{tab:material_params2}
	\setlength{\tabcolsep}{1.3em}
\renewcommand{\arraystretch}{1.8}
	\begin{tabular}{l|l|l|l|l|l|l}
		$c_1$ in MPa & $c_2$ in MPa & $c_3$ in N & $c_4$ in N & $c_5$ in N & $c_6$ in N & $c_7$ in N \\
		\midrule
		6577.18 & 134.23 & 0.59 & 0.59 & 0.18 & -0.23 & 0.18 \\
		\bottomrule
	\end{tabular}
\end{table}
The simple shear problem is investigated for the cases D and T, prescribed displacement and applied traction, respectively. In the case D, $\hat{\ten u}=(0.05, 0)$ in mm is applied for a horizontal displacement of the top edge of the plate. In the second case, a traction vector, $\hat{\ten t}=(1,0)$ in N/mm, applied for a horizontal shear to the same (top) edge. In Fig. \ref{shear_sim}, the numerical predictions obtained by the performed finite element simulations are compared to the analytical solution for both cases. The provided plots are the displacement in the horizontal direction occurring along the right edge of the plate. The zoomed-in sections highlight the predicted displacements and the accuracy of the approximations. Notably, the applied zero displacement gradient conditions in both cases, the top edge for the prescribed displacement case and the bottom edge for the applied traction case, are accurately captured in the finite element simulations.

In general, numerical predictions are consonant with the analytical solution in both cases. Precisely, for the case D, obviously, Argyris and the mixed FE simulations provide a convincing match against the analytical solution, while Hermite and IGA simulations provide slightly different predictions, around the top edge of the plate. On the other hand, for the applied traction case, a similar trend is observed, having negligible differences around the bottom edge of the plate. The discretization of the presented results in Fig.\,\ref{shear_sim} have been selected based on convergence behavior of each formulation. For the simulations with the Argyris and Hermite elements, the rectangular domain is discretized into 54 triangular elements that is 612 Degrees Of Freedom (DOF) and 96 triangular elements that is 552 DOF triangular elements, respectively. A single 2nd-degree NURBS patch with 100 elements (882 DOF) is used for the IGA simulation. And, to perform the mixed FEM simulation, the domain is discretized into 5400 linear elements (54\,720 DOF).

\begin{figure}[H]
	\centering
	\begin{subfigure}{0.49\textwidth}
		\centering
		\includegraphics[width=0.98\linewidth]{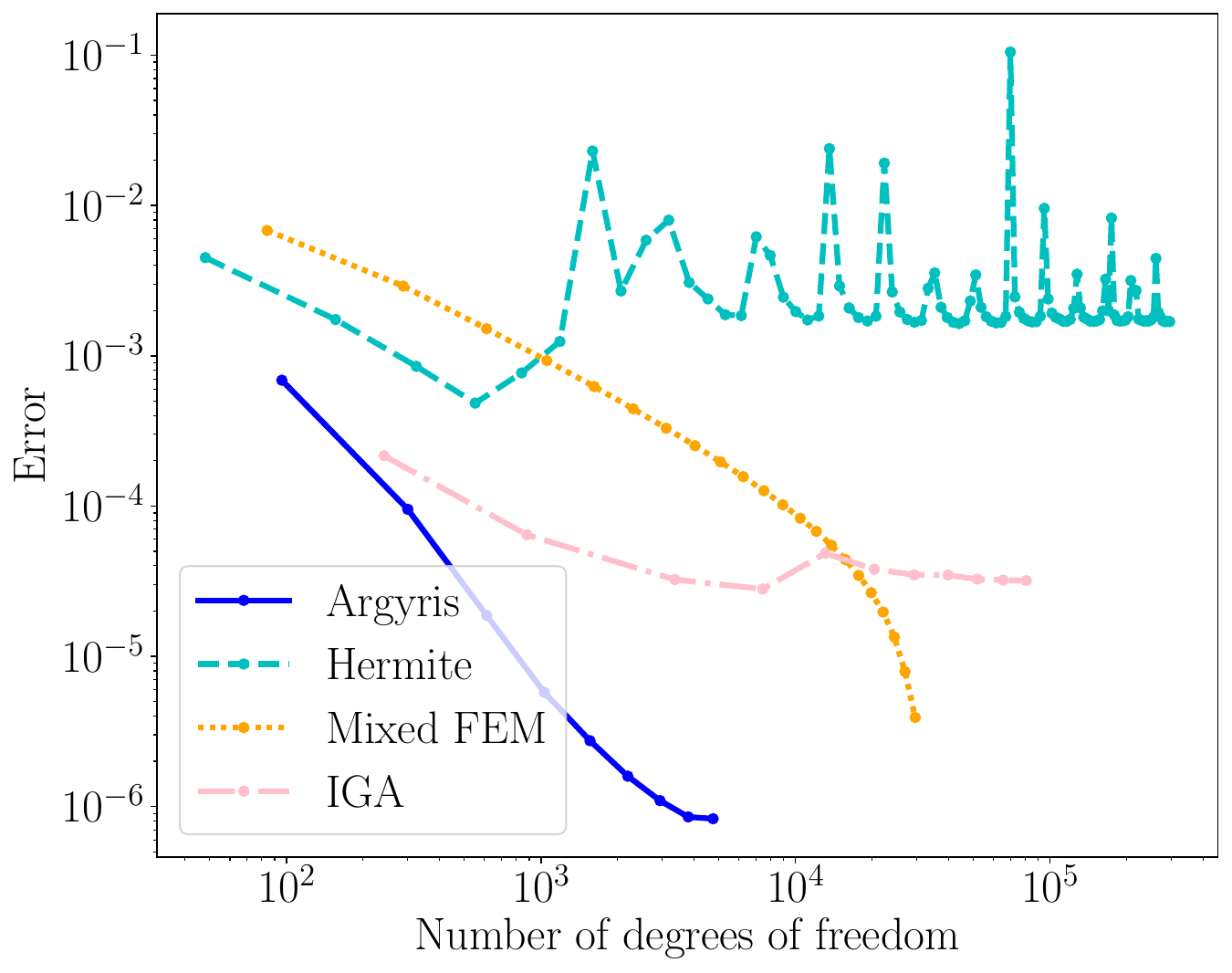}
		\caption{Prescribed displacement case.}
		\label{shear_sim1_error}
	\end{subfigure}
	\begin{subfigure}{0.49\textwidth}
		\centering
		\includegraphics[width=0.98\linewidth]{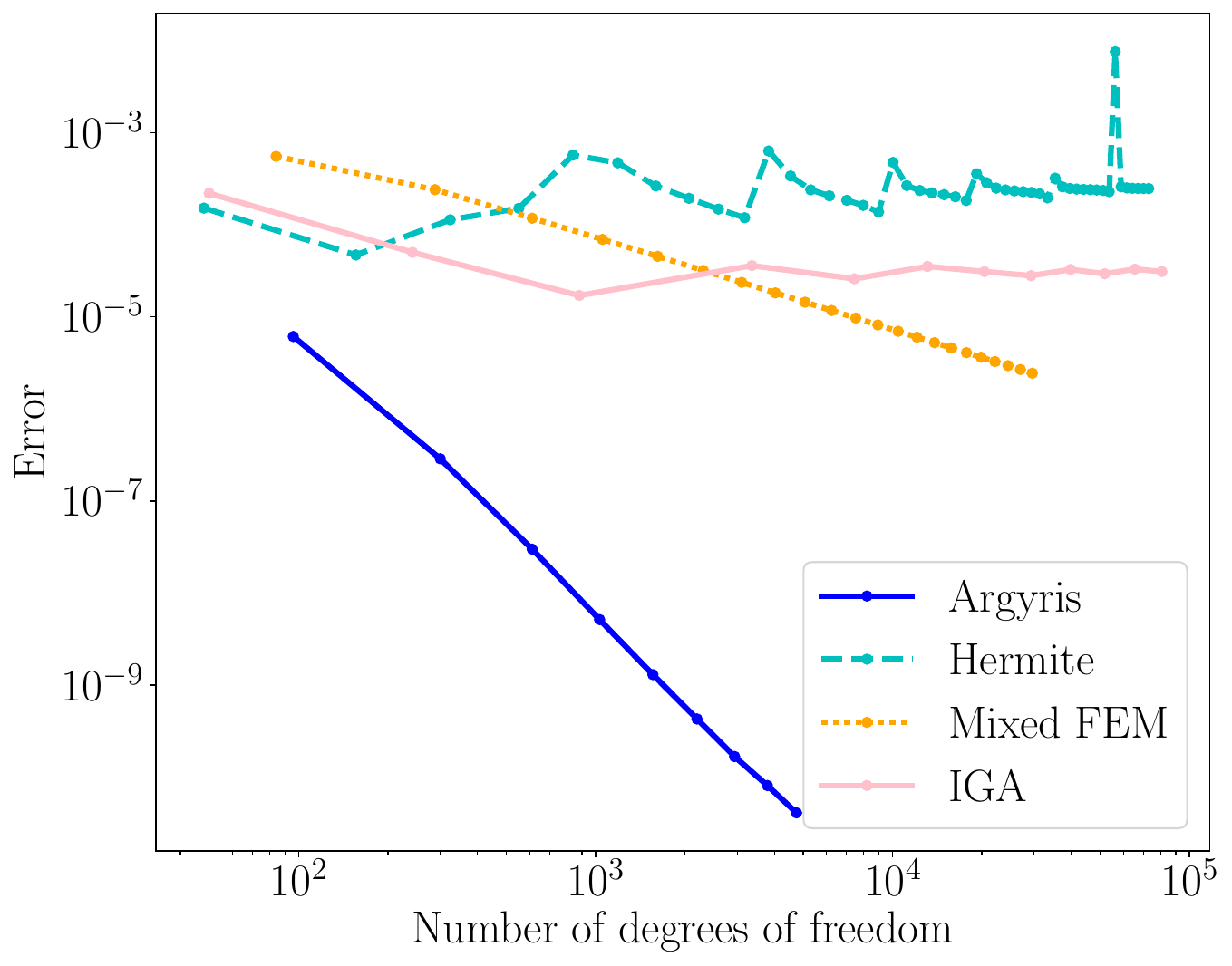}
		\caption{Applied traction case.}
		\label{shear_sim2_error}
	\end{subfigure}
	\caption{Error analysis on a log-log scale for finite elements in the simple shear deformation problem with prescribed displacement and applied traction}
	\label{shear_sim_error}
\end{figure}

Furthermore, to investigate the convergence characteristics of different FE formulations, a series of simulations is designed, refining the computational domain systematically as known as $h$-convergence. The computed errors are presented as a function of degrees of freedom on a log-log scale using $L_1$ error norm for both cases in Fig. \ref{shear_sim_error}, calculated by
\begin{equation}
	\text{Error} = \int_{0}^{1} |u^{\text
	{ref}} - u^{\text{sim}}| \, dx  
	\approx \sum_{i=0}^{N-2} \frac{1}{2} \left( |u^{\text
	{ref}}_i - u^{\text{sim}}_i| + |u^{\text
{ref}}_{i+1} - u^{\text{sim}}_{i+1}| \right) \Delta x
	\label{errornorm}
\end{equation}

using the trapezoidal integration method and considering the analytic solution, $u^{\text
{ref}}$ and the numerical solution, $u^{\text{sim}}$, in 1-D reduced order model. The error calculation involves defining the distance $\Delta x$ as $\frac{1}{N-1}$, where $N$ represents the number of calculated displacement values.

In the simulations with Argyris and the mixed FE, we observe a monotonous convergence. As expected, Argyris elements have a higher convergence rate compared to the mixed FE with an adequate accuracy and efficiency compared with the analytical solution. Remarkably, not ideal convergence characteristics are obtained for Hermite elements and IGA such studied in \cite{riesselmann2024efficient}, with increasing number of degrees of freedom. Especially, the simulations with Hermite elements show an undesirable convergence behavior in both cases such that a posteriori error estimation is not possible without knowing the solution.

An analogous study has been conducted with the 1-D strain-gradient model of pull-out problem. By using different finite element formulations, we model rebar as embedded into a cylindrical concrete block of Young's modulus $E = 20$ GPa and Poisson's ratio $\nu = 0.2$. In Table \ref{tab:material_params1}, all constitutive parameters are calculated by Eq. \eqref{constitutive}. The solution of this geometric nonlinear reduced order problem investigates the accuracy by using a linearization scheme. Herein, we use SNES scheme from PETSc libraries \cite{petsc-web-page}.

\begal
& a_1 = \frac{l^2}{56} \lambda 
\\ 
& a_2 =  \frac{l^2}{224} \lambda
\\ 
& a_3 =  \frac{l^2}{560}(7\mu + 3\lambda) \\ 
& a_4 =  \frac{l^2}{1120}(7\mu - 4\lambda)
\\ 
& a_5 = \frac{l^2}{560}(7\mu + 3\lambda) \ ,
\alend

\begin{table}[H]
	\centering
	\caption{Constitutive parameters for the pull-out problem.}
	\label{tab:material_params1}
		\setlength{\tabcolsep}{1.3em}
\renewcommand{\arraystretch}{1.8}
	\begin{tabular}{c c c c c c c}
		\toprule
		{$\lambda$ (MPa)} & {$\mu$ (MPa)} & {$a_1$ (N)} & {$a_2$ (N)} & {$a_3$ (N)} & {$a_4$ (N)} & {$a_5$ (N)} \\
		\midrule
		5555.55 & 8333.33 & 6.20 & 1.55 & 8.37 & 2.02 & 8.37 \\
		\bottomrule
	\end{tabular}
\end{table}

\begin{figure}[H]
	\centering
	\subfloat[]{
		\includegraphics[width=0.70\textwidth]{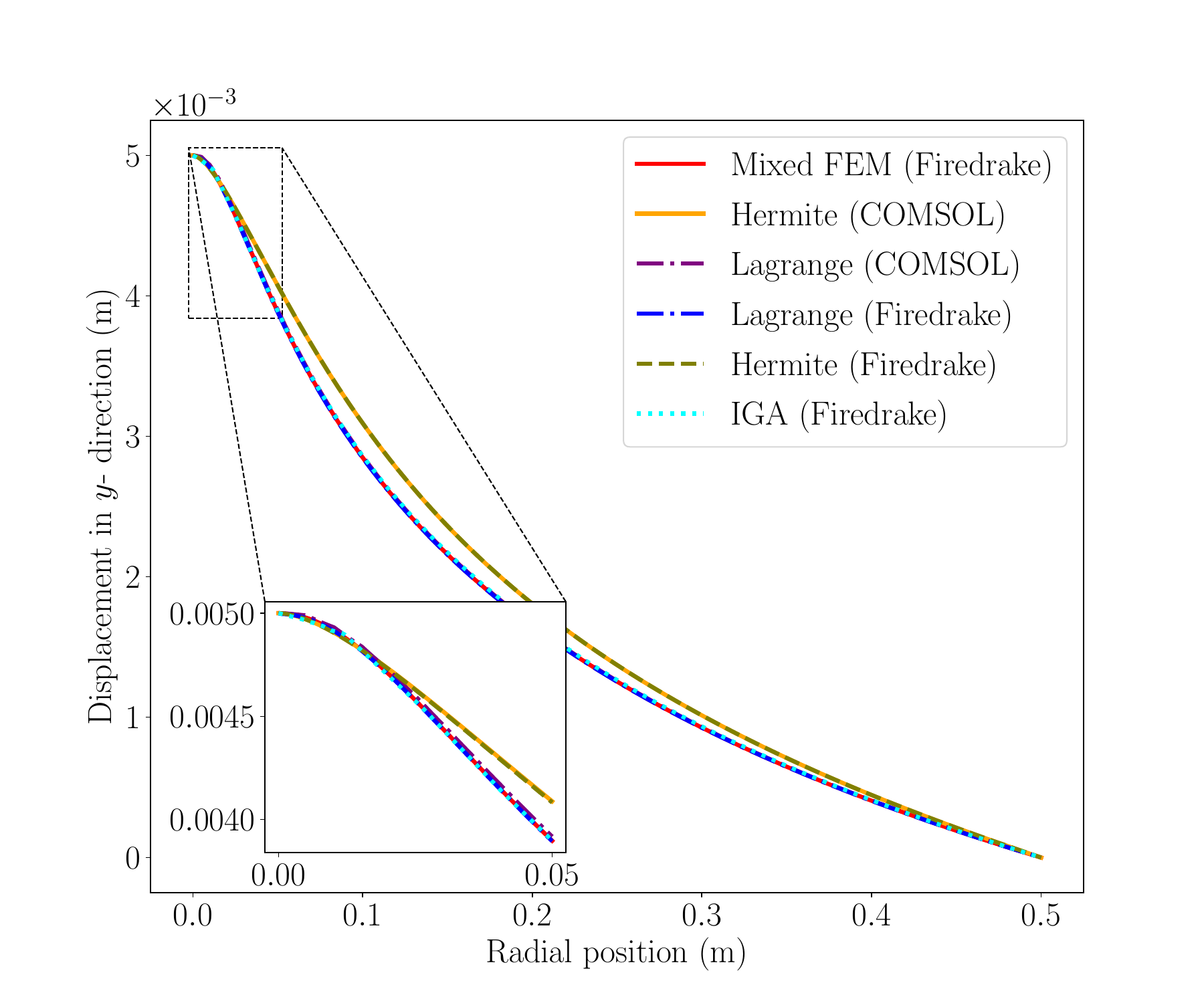}
	}
	\subfloat[]{
		\stackinset{r}{0pt}{b}{+220pt}{
			\includegraphics[width=0.30\textwidth]{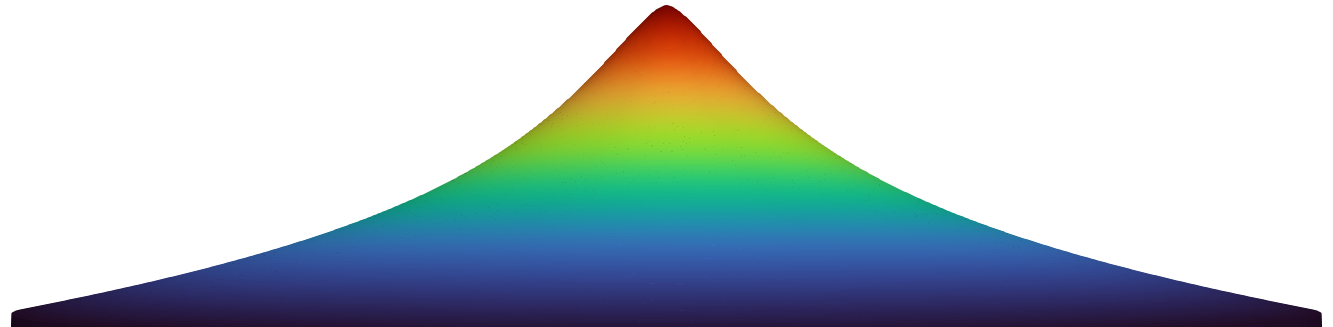}}
		{\includegraphics[width=0.30\textwidth]{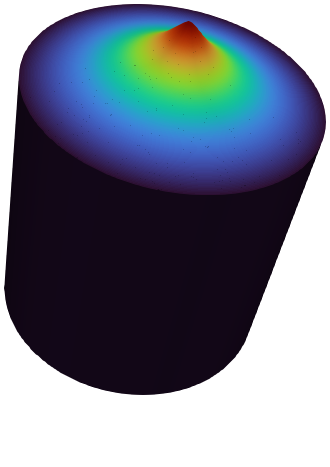}}
	}
	\caption{a) Comparison of the converged results of the strain-gradient elasticity solution of the pull-out problem  \\ b) 3-D illustration of deformation in the pull-out problem, captured in ParaView (scaled 50 times)}
	\label{fig:comp_pull-out}
\end{figure}

The numerical simulations are performed using quadratic Lagrange, cubic Hermite elements, the mixed FE, and IGA with a 2nd-degree NURBS spline. In addition to the simulations on FEniCS and Firedrake, the problem is also investigated by COMSOL Multiphysics\textsuperscript{\textregistered} using cubic Hermite and quadratic Lagrange elements. The weak form of the PDE is implemented by introducing the energy density in COMSOL Multiphysics\textsuperscript{\textregistered}. In Fig.\,\ref{fig:comp_pull-out}, the performed simulations are compared, providing the predicted displacements in the radial direction of the cylindrical block. For each element type, the problem is simulated with 5000 elements on FEniCS and Firedrake. It is clear that numerical simulations with Lagrange element, the mixed FE, and IGA compare each other very well. However, the displacement predicted by Hermite elements is notably different than other FE formulations. The same trend is observed for the simulations performed on COMSOL Multiphysics\textsuperscript{\textregistered}. Overall, the predicted results on FEniCS and Firedrake compare well with those obtained by COMSOL Multiphysics\textsuperscript{\textregistered}.

\begin{figure}[H]
	\centering
	\includegraphics[width=0.5\textwidth]{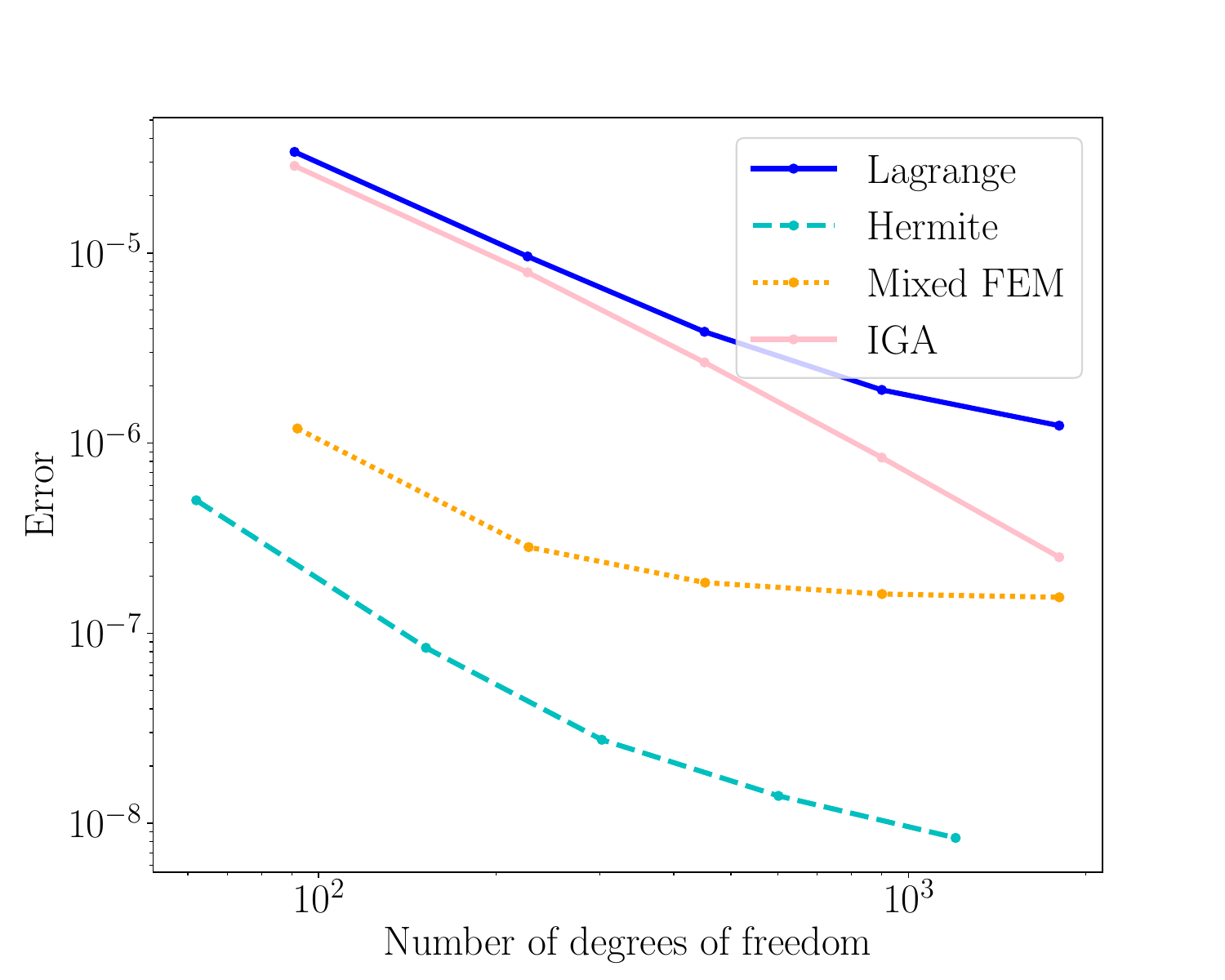}
	\caption{Convergence on a log-log scale for the pull-out problem }
	\label{sim3_error}
\end{figure}

In order to investigate the convergence characteristics, a series of simulations is designed by using four levels of discretization. In this regard, the computational domain is divided into 5, 50, 500, and 5000 elements, respectively, to perform a convergence study. In Fig. \ref{sim3_error}, the convergence of each formulation is presented on a log-log scale using the $L_1$ error norm as defined in \eqref{errornorm}. The most discretized solution, with 5000 elements, is considered the reference solution, $u^{\text{ref}}$. It is clear that monotonic convergence is observed for each element type for the 1-D non-linear strain-gradient model.

Furthermore, CPU-time performance of each finite element formulation is analyzed to assess their computational efficiency. Table \ref{tab:process_time} presents a comparison in terms of CPU-time, considering all the simulations examined in this study. For the 2-D model, it is clear that the mixed FE has the highest CPU-time for both cases, while IGA has the lowest. IGA stands out from other methods due to its efficient assembly process. However, this advantage appears to be more significant in basic problems rather than in complex simulations. For the 1-D model, Lagrange elements have the highest CPU-time, followed by the mixed FE, Hermite elements, and IGA respectively. In this analysis, the number of degrees of freedom is kept constant for each case to ensure a fair comparison. Also, Hermite elements are excluded in the 2-D model due to their convergence behavior.

\begin{table}[H]
		\centering
		\caption{CPU time(Assemble $+$ Computation) performance of different finite element formulations}
		\label{tab:process_time}
		\setlength{\tabcolsep}{1.3em}
		\begin{tabular}{p{2cm}cccccc}
			\toprule
			& \parbox{1.5cm}{\centering \textbf{Element} \\ \textbf{Type}} 
			& \parbox{1.5cm}{\centering \textbf{Number of} \\ \textbf{Nodes}} 
			& \parbox{1.5cm}{\centering \textbf{DOF}}  
			& \multicolumn{2}{c}{\textbf{CPU time (sec)}} \\
			\cmidrule(lr){5-6}
			& & & & \parbox{1.5cm}{\centering \textbf{Assemble} \\ \textbf{Time}} & \parbox{2.5cm}{\centering \textbf{Computation} \\ \textbf{Time}} \\
			\midrule
			\multirow{3}{*}{\parbox{2cm}{\centering \textbf{Case D:} \\ \textbf{Prescribed} \\ \textbf{Displacement}}} 
			& Argyris  & 36   & 612     & 0.123 & 0.438 \\
			& Mixed FEM & 36  & 612     & 0.129 & 0.635\\
			& IGA      & 42   & 608     &  0.122 & 0.062  \\
			\midrule
			\multirow{3}{*}{\parbox{2cm}{\centering \textbf{Case T:} \\ \textbf{Applied} \\ \textbf{Traction}}} 
			& Argyris  & 36   & 612     & 0.124 & 0.451 \\
			& Mixed FEM & 36  & 612     & 0.184 & 0.565 \\
			& IGA      & 42   & 608     &  0.134 & 0.053  \\
			\midrule
			\multirow{4}{*}{\parbox{2cm}{\centering \textbf{Case P:} \\ \textbf{Pull-out} \\ \textbf{Problem}}} 
			& Lagrange  & 36    & 106     & 0.027 & 0.159 \\
			& Hermite   & 53    & 106     & 0.027 & 0.215 \\
			& Mixed FEM & 36    & 107     & 0.037 & 0.318 \\
			& IGA       & 35    & 106     & 0.035  &0.018   \\
			\bottomrule
		\end{tabular}
	\end{table}
	
\section{Conclusion}
In this study, a numerical investigation is presented on the applicability of different finite element formulations within the framework of strain-gradient elasticity by using the open-source platforms FEniCS and Firedrake. To this end, two problems previously examined in the literature are investigated. A 2-D linear strain-gradient model is utilized for the simple shear of a plate. The simulations are performed for two different cases using Argyris, Hermite, a mixed FE, and IGA formulations. It is observed that Argyris and the mixed FE perform well, accurately predicting the problem under study, while Hermite and IGA lack the ideal convergence trends. Furthermore, a 1-D non-linear strain model is applied for the pull-out test, considering a rigid bar embedded into a cylindrical concrete block. For this problem, the 1-D simulations are performed using Lagrange, Hermite, the mixed FE, and IGA formulations. It is observed that all the finite element formulations exhibit monotonic convergence behavior for the 1-D model. It can be emphasized that simulations can be performed in 3D, except for those executed using the Argyris element. Importantly, predicted displacement by Hermite elements is different than those obtained with other formulations. This discrepancy is also confirmed by simulations in COMSOL Multiphysics\textsuperscript{\textregistered} for validation. Moreover, CPU-time performance of each formulation is analyzed to compare their computational efficiency.

\bibliographystyle{unsrt} 
\bibliography{references} 
\end{document}